\documentclass[12pt]{article}

 \newcommand{\HIDDEN}[1]{}
\newcommand{\frakh}{\mathfrak{h}}

\usepackage[dvips]{graphicx}
\usepackage{amsmath}
\usepackage{amsfonts}
\usepackage{amsthm}
\usepackage{placeins}
\usepackage{afterpage}
\usepackage{dsfont}
\usepackage{flafter}     

\usepackage{scrpage}

\usepackage[a4paper]{typearea}
\areaset[0cm]{17.0cm}{27cm}

\newcommand{\PSImagx}[2]{\includegraphics[width=#2]{psplots/#1}} 
\newcommand{\BILD}[4]{\begin{figure}[#1]%

     #2

     \centerline{\parbox{15cm}{\caption[.]{#3} \label{#4}}}
     \end{figure} }

\newcommand{\Int}{\int\limits}

\newcommand{\erfc}{\operatorname{erfc}}

\newcommand{\ud}{\text{d}}
\newcommand{\ui}{\text{i}}
\newcommand{\ue}{\text{e}}

\newcommand{\R}{\mathbb{R}}
\newcommand{\Z}{\mathbb{Z}}
\newcommand{\N}{\mathbb{N}}
\newcommand{\C}{\mathds{C}}

\newcommand{\limacon}{lima\c{c}on }

\newcommand{\setsep}{ \;\; | \;\;}

\newcommand{\cH}{{\cal H}}
\newcommand{\cP}{{\cal P}}

\newcommand{\CL}{{\cal L}}

\newcommand{\cD}{{\cal D}}

\newcommand{\bfA}{\boldsymbol{A}}

\newcommand{\vol}{\operatorname{vol}}

\newcommand{\VoldOmega}{{\vol(\partial \Omega)}}

\newcommand{\Cu}{\overline{u}}

\newcommand{\cs}{c_{(q,p),k}^{b}}
\newcommand{\Ccs}{\overline{c}_{(q,p),k}^{b}}

\newcommand{\csn}{c_{(q,p),k_{n}}^{b}}
\newcommand{\ccsn}{\overline{c}_{(q,p),k_{n}}^{b}}

\newcommand{\pa}{\partial}
\newcommand{\la}{\langle}
\newcommand{\ra}{\rangle}
\newcommand{\e}{\hat {\boldsymbol{e}}}  
\newcommand{\n}{\hat {\boldsymbol{n}}}  
\newcommand{\x}{\boldsymbol{x}}	
\newcommand{\y}{\boldsymbol{y}}
\newcommand{\p}{\boldsymbol{p}}	 
\newcommand{\q}{\boldsymbol{q}}
\newcommand{\z}{\boldsymbol{z}}

\newcommand{\ta}{\hat {\boldsymbol{t}}}

\renewcommand{\Im}{\operatorname{Im}}

\newcommand{\Osk}{O(k^{-1/2})}
\newcommand{\OneOsk}{(1+O(k^{-1/2}))}
\newcommand{\OneOskn}{(1+O(k_{n}^{-1/2}))}
\newcommand{\OneOk}{(1+O(k^{-1}))}
\newcommand{\Ok}{O(k^{-1})}

\begin{document}



\newcommand{\titel}{Poincar\'e Husimi representation\\[1ex]
of eigenstates in quantum billiards}

\normalsize

\vspace{2.5cm}

\begin{center}  \huge  \bf
      \titel

\end{center}

\vspace{1.5cm}

\begin{center}

         {\large A.\ B\"acker%
\footnote{E-mail address: {\tt baecker@physik.tu-dresden.de}}$^{)}$, 
                 S.\ F\"urstberger%
\footnote{E-mail address: {\tt silke.fuerstberger@physik.uni-ulm.de}}$^{)}$, 
                 R.\ Schubert%
\footnote{E-mail address: {\tt schubert@msri.org}}$^{)}$
         }

   \vspace{4ex}

  \begin{minipage}{10.75cm}

   \begin{tabular}{ll}
   1)\hspace*{-0.25cm}& Institut f{\"u}r Theoretische Physik,
   TU Dresden,\\
   &D--01062 Dresden, 
   Germany\\[2ex]
   2)\hspace*{-0.25cm}&Abteilung Theoretische Physik, Universit\"at Ulm\\
   &Albert-Einstein-Allee 11, D--89069 Ulm, Germany\\[2ex]
   3)\hspace*{-0.25cm}&
Department of Mathematics, University of Bristol \\
& University Walk, Bristol, BS8 1TW, UK 
\\[2ex]
  \end{tabular}
  \end{minipage}

\end{center}

\renewcommand{\thefootnote}{\arabic{footnote}}
\setcounter{footnote}{0}


\vspace*{1.5cm}
 
\leftline{\bf Abstract:} 

\noindent
For the representation of eigenstates 
on a Poincar\'e section at the boundary of a billiard
different variants have been proposed.
We compare these Poincar\'e Husimi functions,
discuss their properties and based on this select one particularly
suited definition.
For the mean behaviour of these Poincar\'e Husimi functions
an asymptotic expression is derived, including
a uniform approximation.
We establish the relation
between the Poincar\'e Husimi functions and
the Husimi function in phase space
from which a direct physical interpretation follows.
Using this, a quantum ergodicity theorem for the Poincar\'e Husimi functions
in the case of ergodic systems is shown.

\newpage

\section{Introduction}

The study of eigenfunctions of quantum systems, in particular
their dependence on the  
classical dynamics, has attracted a lot of attention.
A prominent class of examples is provided by
two--dimensional billiard systems, which are classically given
by the free motion of a particle inside some domain with
elastic reflections at the boundary.
The corresponding quantum system is described by the
Helmholtz equation
inside a compact domain $\Omega \subset \R^2$ 
(in units $\hbar=1=2m$),
\begin{equation} \label{eq:Helmholtz}
  \Delta \psi_n(\x) + k_n^2 \psi_n(\x) = 0 \;\; , 
            \qquad \x \in \Omega  \;\;,
\end{equation}                                                         
with (for example) Dirichlet boundary conditions 
\begin{equation} 
  \label{eq:DC-bc}
  \psi_n(\x) = 0  \;\; , \qquad \x \in \partial \Omega \;\;,
\end{equation}
where the eigenfunctions $\psi_n(\x)$ are in $L^2(\Omega)$.
Assuming that the eigenvalues $k_n^2$ are ordered with increasing value,
the semiclassical limit corresponds to $n\to\infty$.
A detailed knowledge of the behaviour of the eigenvalues $k_n^2$
and the structure of eigenstates is relevant for applications,
for example microwave cavities 
or mesoscopic systems (see e.g.~\cite{Sto99} and references therein).

For the description of the phase space structure of
quantum systems usually the Wigner function \cite{Wig32} or
Husimi function \cite{Hus40} are used.
However, for a system with $d$ degrees of freedom
these are $2d$--dimensional functions, which are difficult to visualize
for $d>1$. Therefore, one usually considers
the position representation, or the momentum representation \cite{BaeSch99},
or sections through the Wigner or Husimi function, see e.g.~\cite{VebRobLiu99}.

Another approach is the use of representations on the billiard
boundary, acting as a global Poincar\'e section. 
In the literature one can find several variants for 
these representations,
see e.g.\ \cite{CrePerCha93,TuaVor95,SimVerSar97}.
The reason is, as emphasized in \cite{TuaVor95},
that there is no natural definition of a scalar product
for functions on the billiard boundary.
This raises the question whether one of these definitions
has advantages over the others, which will be addressed in the following.

The representation of eigenstates on the Poincar\'e section plays
an important role in several applications. For example, it is
used to define scar measures \cite{SimVerSar97,BieKapHagHel2001},
or to study  conductance fluctuations, see \cite{Kap2000} and
references therein. Furthermore, these representations are used
to determine the coupling of leads in open systems
\cite{BaeManHucKet2002}.
Another important application is the
detection of regions where eigenstates localize, see e.g.
\cite{BaeSch2002a,BaeSch2002b,BaeManHucKet2002}
(for an alternative approach based on the scattering approach see
\cite{KlaSmi96,FriDor97}).
Representations of eigenstates on the Poincar\'e section
have also been useful to understand the behaviour of 
optical microresonators, see e.g.~\cite{HenSchSch2003}
and references therein.
More generally, the approach is not just applicable for 
billiard systems but it is also useful for
Poincar\'e sections arising from Bogomolny's transfer
operator approach \cite{Bog92}.

In this paper we first compare two different definitions for the Poincar\'e
Husimi representation,
discuss their properties (section \ref{sec:hus-defs}) 
and based on this we select one particular definition
for the following.
In section~\ref{sec:mean-behaviour} we derive the behaviour 
of these Poincar\'e Husimi functions when averaged
over several energies. 
In section~\ref{sec:Poincare-Husimi-vs-Husimi}
we establish a relation between the well--known Husimi function
in phase space and the Poincar\'e Husimi function on the billiard
boundary. This allows for a direct physical interpretation of
the Poincar\'e Husimi functions. 
Moreover, for ergodic systems a 
quantum ergodicity theorem for the Poincar\'e Husimi functions
is shown.

\section{Husimi representation on the boundary} \label{sec:hus-defs}

Let us first recall the definition and some properties of Husimi functions
in phase space. 
For a solution $\psi_n$ of the Helmholtz equation 
\eqref{eq:Helmholtz} with energy $E=k_n^2$
the Husimi function $H_n^B(\p,\q)$ 
is given by its projection onto a coherent state, i.e.\
\begin{equation}\label{eq:def-of-int-Hus}
H^B_n(\p,\q):= 
\left(\frac{k_n}{2\pi}\right)^2 
     \left|\la \psi^{B}_{(\p, \q),k_n},\psi_n\ra_{\Omega}\right|^2\, \,. 
\end{equation}
Here
\begin{equation}
  \la \psi_1,\psi_2\ra_{\Omega}
 := \Int_\Omega \overline{\psi}_1(\q)\psi_2(\q)\, \ud^2 q
\end{equation}
is the scalar product in $\Omega$,
and $\overline{\psi_1}$ denotes the complex conjugate of $\psi_1$.

The coherent states are defined as 
\begin{equation}\label{eq:def-of-int-cs}
\psi^{B}_{(\p,\q),k}(\x):=\Bigl(\frac{k}{\pi}\Bigr)^{1/2}\, (\det \Im B)^{1/2}
\ue^{\ui k[\la \p ,\x-\q\ra+\frac{1}{2}\la \x-\q, B(\x-\q)\ra]}\;\;,
\end{equation}
where $(\p, \q)\in \R^2\times\R^2$ denotes the point in phase space around 
which the coherent state is localized, and 
$B$ is a symmetric complex $2\times 2$ matrix 
which determines the shape of the coherent state.
For the conventional coherent states one has $B=\ui\, 
\bigl(\begin{smallmatrix} 1&0\\ 0&1\end{smallmatrix}\bigr)$
an in general one has the condition $\Im B>0$,
i.e.\  $\langle v,\Im B \,v\rangle >0$ for all $v \in R^2\backslash\{0\}$.
Notice, that because the variance of the coherent states is proportional  
to $k$, all Husimi functions are concentrated around the energy shell 
$|\p|^2=1$ (and not around $|\p|^2=k^2$).
By this it is possible
to compare Husimi functions with different energies $k_n^2$, and for example
consider their mean, see \eqref{eq:int-mean-value} below.

Such Husimi functions can be  interpreted as probability distributions
on phase space, because they satisfy the relation 
\begin{equation}\label{eq:int-prob-distr}
\la \psi_n, \bfA \psi_n\ra_{\Omega} =\Int_{\R^2}\! \Int_{\R^2}a(\p,\q)
	H^B_n(\p,\q)\, \ud^2 p\,\ud^2 q+\Ok
\end{equation}
where $a(\p,\q)$ is a function on phase space and $\bfA$ its quantization. 
Moreover, the average of all Husimi functions
$H^B_n(\p,\q)$ up to some energy $k^2=E$ 
converges for $k\to\infty$ 
to the normalized invariant measure on the energy shell, 
\begin{equation}\label{eq:int-mean-value}
\lim_{k\to\infty}\frac{1}{N(k)}\sum_{k_n\leq k}H^B_n(\p,\q)=
\frac{1}{\pi\vol{(\Omega})}\, \chi_{\Omega}(\q)\delta(1-|\p|^2)\,\, .
\end{equation}   
Here $N(k)$ denotes the spectral staircase function,
$N(k):=\#\{k_n \leq k \}$, and $\chi_{\Omega}(\q)$ is the characteristic 
function on $\Omega$.
The mean behaviour \eqref{eq:int-mean-value}
is similar to the mean behaviour of the spectral staircase function,
which is given by the Weyl formula,
i.e.\ for $k\to\infty$ one has
     $N(k) \sim \frac{A}{4\pi} k^2$
where $A$ is the area of the billiard.
A similar asymptotic behaviour can be derived for the
mean of normal derivative functions, see \cite{BaeFueSchSte2002}
for a detailed discussion.

For billiards an extremely useful approach 
for describing the dynamics is the use of a Poincar\'e 
section $\cP$  together with the corresponding Poincar\'e mapping $P$.
Usually the section 
$\cP := \{ (q,p) \setsep q\in [0,\VoldOmega],\;  p \in [-1,1] \}$ is 
parameterized by the arc-length coordinate $q$ along the boundary 
$\partial\Omega$ of the billiard
and the projection $p$ of the (unit) momentum $\hat\p$
after the reflection on the tangent $\ta(q)$, 
i.e.\ $p=\la\,\hat\p,\ta(q)\ra$.
By this the billiard flow induces an area--preserving
map $P:\cP\to\cP$ where
the invariant measure is given by
$\ud \mu = \ud q \,\ud p$.

In order to have the advantages of such a reduced representation 
in quantum mechanics as well, one is 
interested in a Husimi representation $h_n(q,p)$ on the
Poincar\'e section $\cP$ which is associated with an eigenstate
$\psi_n$. Such a {\it Poincar\'e Husimi function}
 should have similar pro\-perties 
as the ones expressed by relations \eqref{eq:int-prob-distr} and 
\eqref{eq:int-mean-value} for the Husimi functions in phase space, and 
our aim is to study to what extent this is possible. 
More precisely, one would like that for the Husimi function on the billiard
boundary a spectral average
\begin{equation} \label{eq:mean-boundary-husimi-def}
   \cH_k(q,p) := \frac{1}{N(k)} \sum_{k_n \le k}  h_n(q,p)
\end{equation}
tends to the invariant measure on $\cP$ as $k\to\infty$,
in the same way as in \eqref{eq:int-mean-value}.

\BILD{t}
     {
     \begin{center}
        \PSImagx{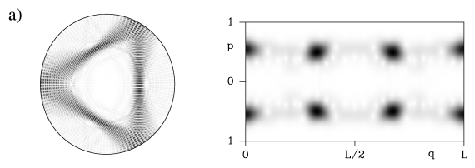}{14cm}\vspace*{0.5cm}

        \PSImagx{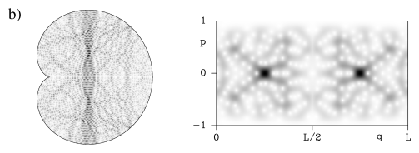}{14cm}\vspace*{0.5cm}

        \PSImagx{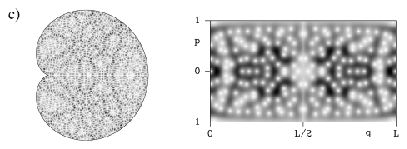}{14cm}\vspace*{0.5cm}
     \end{center}
     }
     {Examples of eigenstates $\psi_n(\q)$, shown to the left,
      and to the right their Poincar\'e Husimi functions $h_n(q,p)$.
      In a) an eigenstate ($n=1952$) localizing around a regular orbit 
      for the \limacon billiard at $\varepsilon=0.3$ is shown.
      In b) and c) two eigenstates for the cardioid billiard
      are shown ($n=1817$ and $n=1277$).
     }
     {fig:husimi-example}

\newcommand{\einesumme}[1]{\PSImagx{#1}{13.75cm}}

\BILD{tbh}
     {
     \begin{center}
      \vspace*{-1cm}
          \einesumme{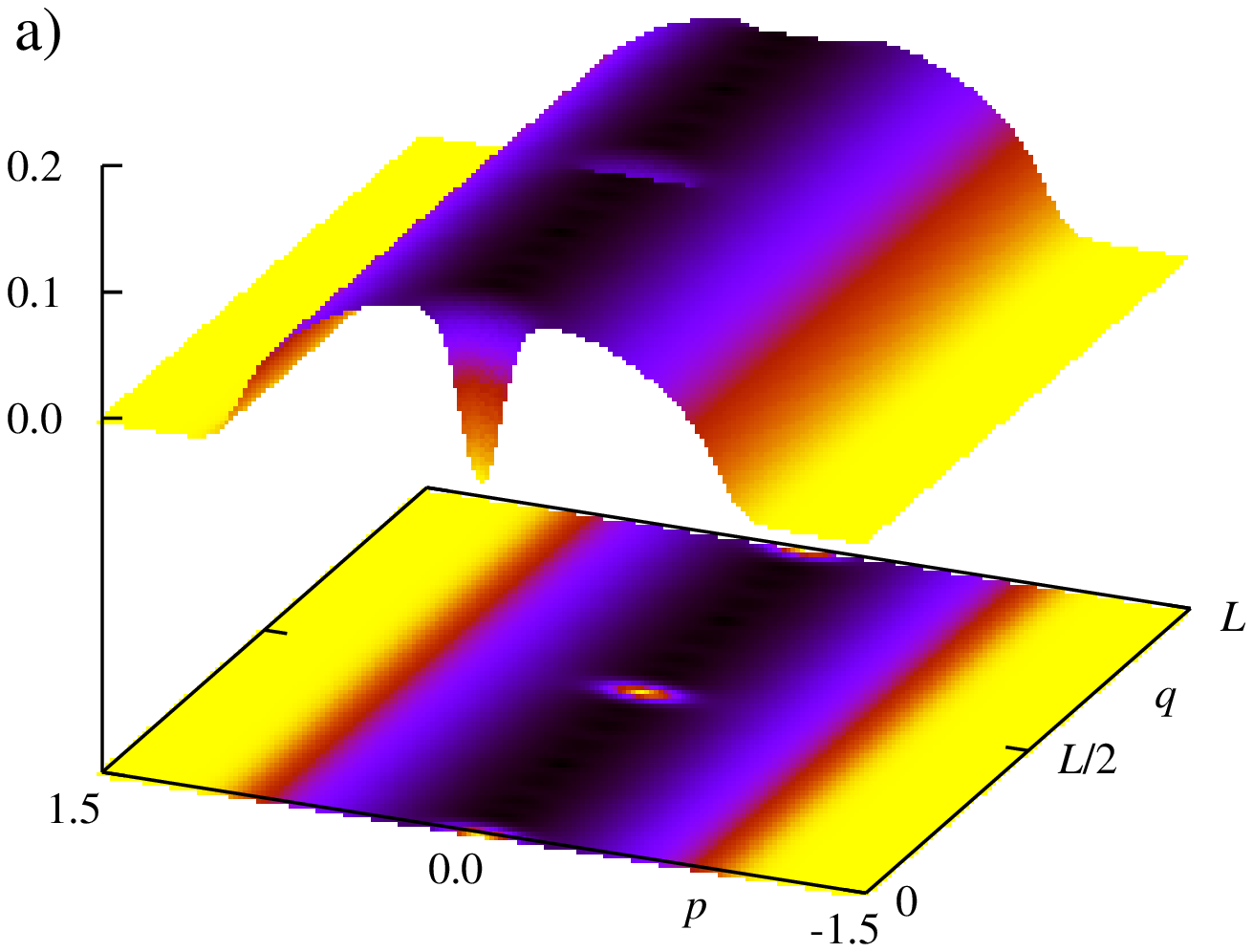}   

\vspace*{-0.25cm}

         \einesumme{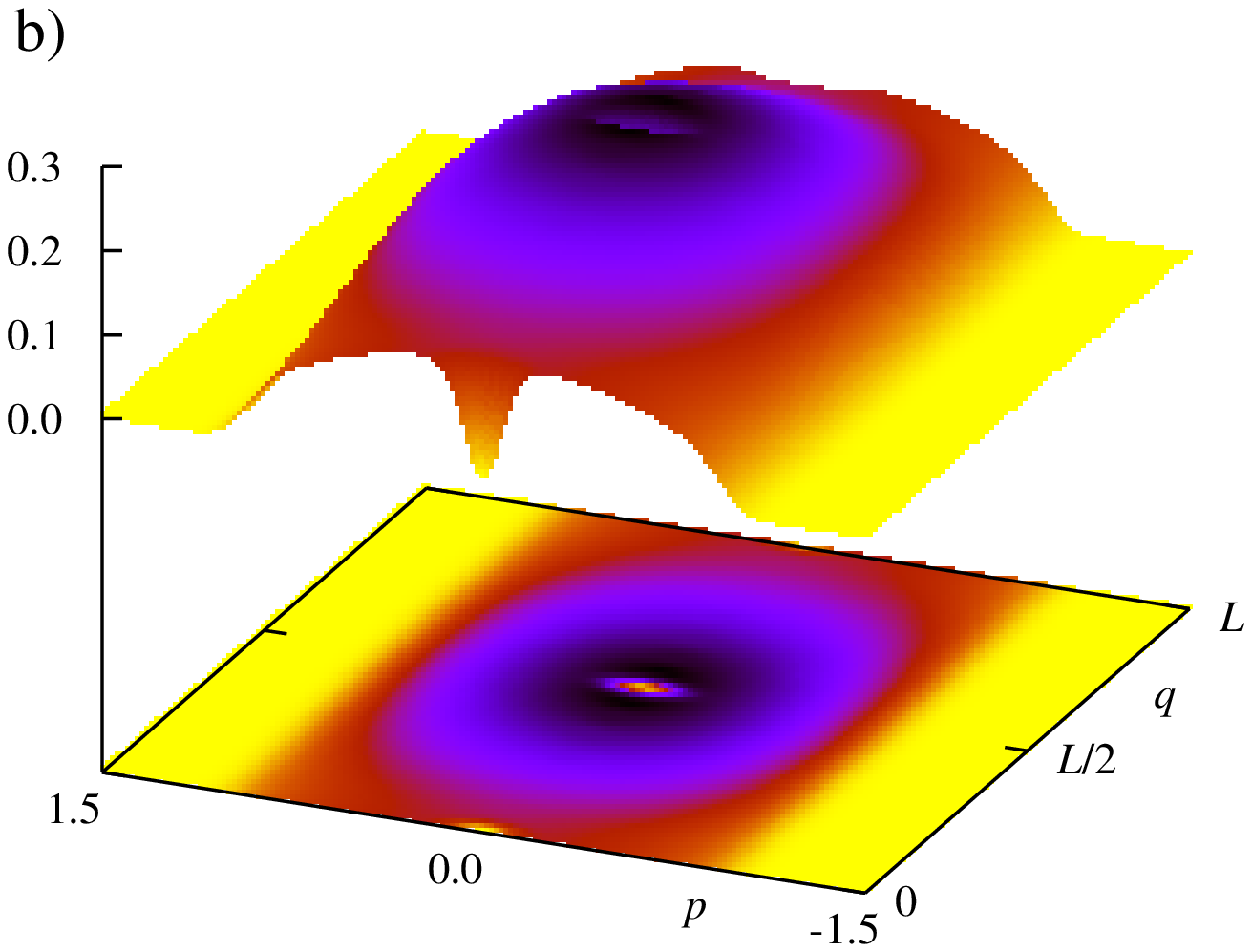}   
\vspace*{-0.5cm}
     \end{center}
     }
     {Plot of $\cH_k(q,p)$ for $k=125$ using the first 2000 eigenstates
      in the \limacon billiard of odd symmetry at $\varepsilon=0.3$.
     In a) the result for $\cH_k(q,p)$
     using definition \eqref{eq:def-p-husimi-0} for $h_n(q,p)$
     is shown and in b) a corresponding $\widetilde{\cH}_k(q,p)$ 
     using definition 
     \eqref{eq:def-p-husimi-2-SimVerSar} is displayed.
     In addition to the symmetry related dips at $(q,p)=(0,0)$ and
     $(L/2,0)$ one clearly sees the variation in $p$--direction in both cases
     and in b) we moreover observe a variation in $q$.
     }
     {fig:sandburgen}

The Husimi representation on the billiard boundary is usually defined 
using the normal derivative of the eigenfunction 
(hereafter called the boundary function)
\begin{equation}\label{eq:def-of-normal-der}
  u_n(s) : = \la\n(s), \nabla \psi_n(\x(s))\ra\,\, ,
\end{equation}
where $\x(s)$ is a point on the boundary $\pa \Omega$, 
parameterized by the arc-length $s$ and $\n(s)$ denotes the outer 
normal unit vector to $\pa \Omega$ at $\x(s)$. 
The boundary functions are 
a natural starting point for defining a Husimi representation because 
they determine the eigenfunctions uniquely, 
see \eqref{eq:u-to-psi}. Thus the boundary functions form a reduced 
representation of the system.
If an eigenfunction $\psi_n$ is normalized,
then the corresponding boundary function $u_n$
fulfils the  normalization condition \cite{Rel1940} 
\begin{equation} \label{eq:u-normalization}
  \frac{1}{2} \Int_{\partial \Omega} 
    | u_n(s) |^2 \,\la \n(s),\x(s)\ra \,\ud s = k_n^2 \;\;.
\end{equation}
For alternative derivations of 
\eqref{eq:u-normalization} and 
more general boundary conditions see 
\cite{BerWil84,Boa94}.
Notice that while the integrand depends on the 
chosen origin for the vector $\x(s)$, the integral  
is independent of this choice.

Starting from the boundary function a Husimi function on the Poincar\'e
section can be defined by a projection onto a coherent state.
There are different possibilities to define coherent states on the 
boundary of a billiard. A
natural choice is the 
periodization of the usual one--dimensional coherent states,
\begin{equation}\label{eq:def-of-bound-cs}
\cs(s):=\Bigl(\frac{k}{\pi}\Bigr)^{1/4}
(\Im b)^{1/4}\sum_{m\in\Z}
        \ue^{\ui k[p(s-q+mL)+\frac{b}{2}(s-q+mL)^2]}\,\, ,
\end{equation}
where  $(q,p)\in  \partial\Omega\times\R$, 
and $L$ denotes the length of the 
boundary. The  parameter $b\in\C$, $\Im b>0$,
determines the shape of the coherent state.
Then for an eigenstate $\psi_n$ with boundary function $u_n$
a Husimi function on the 
Poincar\'e section  $\cP$
(or more precisely, on the cylindric
phase space $\pa\Omega\times\R$) 
can be defined as \cite{CrePerCha93,TuaVor95}
\begin{equation} \label{eq:def-p-husimi-0}
 h_n(q,p) =  
      \frac{1}{2\pi k_{n}}
      \left|\;\;
         \Int_{\partial\Omega}\ccsn(s) \; u_n(s)  \; \ud s 
      \right|^2 \;\;.
\end{equation}
The completeness relation for the coherent states gives
\begin{equation}
\Int_{\partial\Omega}\Int_{\R} h_n(q,p)\, \ud p\,\ud q =
\frac{1}{k_n^2}\Int_{\partial\Omega} |u_n(s)|^2\, \ud s\,\, ,
\end{equation}
so in view of relation
\eqref{eq:u-normalization} the Poincar\'e Husimi function
$h_n(q,p)$ will in general 
not be normalized. This can be fixed by dividing $h_n(q,p)$ by  the 
factor 
$\frac{1}{k_n^2}\int |u_n(s)|^2\, \ud s$, 
as was done for instance in \cite{BaeSch2002a,BaeSch2002b}.
But later on we will see that it is more natural to work with the 
non--normalized Husimi functions \eqref{eq:def-p-husimi-0}.

A different Poincar\'e representation has been proposed in \cite{SimVerSar97},
\begin{equation}  \label{eq:def-p-husimi-2-SimVerSar}
 \tilde{h}_n(q,p) =  
       \frac{1}{2k_n^2} 
       \frac{\left| \Int_{\partial\Omega} \ccsn(s) \; u_n(s)  
                       \; \la\n(s),\x(s)\ra\; \ud s \right|^2}
                  {\Int_{\partial\Omega}   
                              \ccsn(s) \csn(s) 
                            \;   \la\n(s),\x(s)\ra\; \ud s }     \;\;,
\end{equation}
where the inclusion of the factor $\la\n(s),\x(s)\ra$ is motivated
by its appearance in the normalization condition  \eqref{eq:u-normalization}.
In order to compare the two definitions, we use the fact that for large $k$ 
the coherent state becomes more and more 
 concentrated around $s=q$ and so 
$\la\n(s),\x(s)\ra\,\ccsn(s)\sim\la\n(q),\x(q)\ra\,\ccsn(s)$.
This 
leads to the relation 
\begin{equation}\label{eq:husimi-comparison}
\tilde{h}_n(q,p)\sim\la\n(q),\x(q)\ra \, h_n(q,p)\,\, 
\end{equation}
between the two definitions for Husimi functions.

Let us first illustrate the behaviour of the 
Husimi representation  given 
by \eqref{eq:def-p-husimi-0}.
As a concrete example we consider a member of the family of 
\limacon billiards introduced by Robnik \cite{Rob83,Rob84},
whose boundary is given in polar coordinates by 
$\rho(\varphi)=1+\varepsilon \cos(\varphi)$ where $\varepsilon\in[0,1]$
is the family parameter. At $\varepsilon=0.3$ the billiard has
a mixed phase space (see figure~1 in~\cite{BaeSch2002a})
and at $\varepsilon=1$ it turns into the fully chaotic
(i.e.\ ergodic, mixing, \dots) cardioid billiard.
Because of the symmetry of the billiard we consider
the half \limacon billiard with Dirichlet boundary conditions
everywhere.
Fig.~\ref{fig:husimi-example} shows a comparison
of eigenstates $\psi_n(\q)$ with their Husimi representations $h_n(q,p)$
as grey-scale plots with black corresponding to large values.
For the computations \mbox{$b:=\ui\sigma^{-1}=\ui$} was chosen.
In a) an eigenstate which is localized around a stable periodic orbit
with period three is shown which is clearly reflected
in its Poincar\'e Husimi function to the right.
The symmetry $h_n(q,p)=h_n(q,-p)$ is due to the time--reversal
symmetry of the system
and the symmetry $h_n(q,p)=h_n(L-q,p)$ stems
from the reflection symmetry of the system.
The plots in fig.~\ref{fig:husimi-example}b) and c)
are at $\varepsilon=1.0$, i.e.\ for the cardioid billiard.
The eigenstate shown in b) 
is localized around an unstable periodic orbit of period two
which is also nicely seen in the prominent peaks for the
corresponding Poincar\'e Husimi function.
In c) an irregular state in the cardioid billiard is displayed which 
is spread out over the full billiard and
also $h_n(q,p)$ does not show any prominent localization.

Now we turn to a comparison of the two
Poincar\'e Husimi representations given 
by \eqref{eq:def-p-husimi-0} and \eqref{eq:def-p-husimi-2-SimVerSar}.
In figure~\ref{fig:sandburgen}
a plot of $\cH_k(q,p)$ 
is shown where $k=125.27\dots$ is chosen such that the first 2000 states
are taken into account.
Both definitions, equations~\eqref{eq:def-p-husimi-0} 
and \eqref{eq:def-p-husimi-2-SimVerSar},
lead to a similar non--uniform behaviour of $\cH_k(q,p)$
in $p$ direction. We will discuss this
behaviour in more detail in the following section.
In addition we observe that $\cH_k(q,p)$ has a minimum 
at $(q,p)=(0,0)$ and $(q,p)=(\pm \CL/2,0)$
which is due to the desymmetrization.
Figure~\ref{fig:sandburgen}b)  shows a plot 
of $\tilde{\cH}_k(q,p)$ which is defined
as $\cH_k(q,p)$, but instead of $h_n(q,p)$ the functions $\tilde{h}_n(q,p)$
are used, see definition \eqref{eq:def-p-husimi-2-SimVerSar}.
In this case we observe in addition a clear variation in $q$.
The reason for this is the factor $\la\n(q),\x(q)\ra$ as 
explained by relation \eqref{eq:husimi-comparison}.
Another important point is that the definition
\eqref{eq:def-p-husimi-2-SimVerSar} depends on the chosen origin 
as the factor $\la\n(q),\x(q)\ra$ does, 
and therefore the integrals in equation~\eqref{eq:def-p-husimi-2-SimVerSar}
are not invariant under a shift of the origin.
Because of the variation of $\tilde{h}_n(q,p)$ in $q$ and the dependence
on the origin we prefer the definition \eqref{eq:def-p-husimi-0}
and will use this exclusively in the following.

\section{Mean behaviour of boundary Husimi functions}\label{sec:mean-behaviour}

In this section we determine the asymptotic behaviour of
the mean $\cH_k(q,p)$
of the boundary Husimi functions for large energies.
To this end we will use the methods from our previous work 
\cite{BaeFueSchSte2002}. Let us introduce
\begin{equation}
g^{\rho}(k,s,s'):=\sum_{n\in \N}\frac{u_n(s) \Cu_n(s')}{k_n^2}
\rho(k-k_n)\, \, ,
\end{equation}
where $\rho$ is a smooth function whose Fourier transform 
is supported in a neighbourhood $[-\eta,\eta]$ 
with $\eta$ smaller than the length of the
shortest periodic orbit of the 
billiard flow. The function $g^{\rho}(k,s,s')$ 
was studied in \cite{BaeFueSchSte2002} and 
an asymptotic expansion was derived.
Its leading term reads 
\begin{equation}\label{eq:g-rho}
g^{\rho}(k,s,s')=\frac{k}{2\pi^2} \Int_0^{2\pi} 
\la \n(s),\e(\varphi)\ra\la \n(s'),\e(\varphi)\ra
\ue^{\ui k\la  \x(s)-\x(s'),\e(\varphi)\ra}\,\, \ud \varphi\; \OneOk\,\, ,
\end{equation}
where $\x(s)$ denotes the position vector on the boundary at point $s$, 
$\n(s)$ denotes the outer unit normal vector to the boundary at $s$
and $\e(\varphi)=(\cos\varphi,\sin\varphi)$ 
is the unit vector in direction $\varphi$.

Multiplying \eqref{eq:g-rho} with $\Ccs(s)$ and $\cs(s')$ and 
integrating over $s$ and $s'$ leads to 
\begin{equation}\label{eq:mult-by-coh-states}
\begin{split}
\sum_{n\in \N} &\rho(k-k_n)h_n(q,p)\\
&=\frac{k^2}{4\pi^3}\Int_0^{2\pi} 
\bigg|\Int_{\pa\Omega}\la \n(s),\e(\varphi)\ra \,
\ue^{\ui k\la  \x(s),\e(\varphi)\ra}\,\Ccs(s)\,\ud s\bigg|^2\,  \ud\varphi 
\;\OneOk\,\, .
\end{split}
\end{equation}
The $s$--integral can be computed by the method of stationary phase, 
\begin{equation}
\begin{split}
\Int_{\pa \Omega} \la \n(s), &  \e(\varphi)\ra 
   \ue^{\ui k\la \x(s),\e(\varphi)\ra}\,\Ccs(s)\, \ud s\\
&=\Bigl(\frac{k}{\pi}\Bigr)^{1/4}(\Im\, b)^{1/4}
\Int_{-\infty}^{\infty}\la \n(s),\e(\varphi)\ra\,
\ue^{\ui k[\la \x(s),\e(\varphi)\ra-p(s-q)-\frac{\bar b}{2}(s-q)^2]}\, \ud s\\
&=\Bigl(\frac{4\pi}{k}\Bigr)^{1/4}
\frac{(\Im\, b)^{1/4}}{[\ui\tilde{b}]^{1/2}}\,\la \n(q),\e(\varphi)\ra\, 
\ue^{\ui k[\la \x(q),\e(\varphi)\ra
     +\frac{1}{2\tilde{b}}(p-\la \ta(q),\e(\varphi)\ra)^2]}\;\OneOsk\,\, ,
\end{split}
\end{equation}
with
\begin{equation} \label{eq:tilde-b-def}
  \tilde{b}=\bar b+\kappa(q)\la \n(q),\e(\varphi)\ra \;\;,
\end{equation}
where $\kappa(q)$ is the curvature of the boundary at $q$.
Inserting this result we obtain 
\begin{equation}\label{eq:mean-int}
\begin{split}
 \sum_{n\in \N} &\rho(k-k_n)h_n(q,p)\\
&=\frac{2k^2}{(2\pi)^3}\Bigl(\frac{4\pi}{k}\Bigr)^{\frac{1}{2}}
\Int_0^{2\pi}\frac{(\Im b)^{1/2}}{|\tilde{b}|}\, |\la\n(q),\e(\varphi)\ra|^2\,
\ue^{-k\frac{\Im b}{|\tilde{b}|^2}
       (p-\la \ta(q),\e(\varphi)\ra)^2}\, \ud \varphi\,  
   \OneOsk\, ,
\end{split}
\end{equation}
and for $|p|<1$ the $\varphi$--integral can again be solved by the method of
stationary phase (notice that there are two stationary points) 
which yields
\begin{equation}\label{eq:mean-beh}
 \sum_{n\in \N}\rho(k-k_n)h_n(q,p)
=\frac{k}{\pi^2 }
\sqrt{1-p^2} \; \OneOsk \,\, .
\end{equation}
By integrating this equation, and using a delta sequence for $\rho$ as in 
proofs of the Weyl formula (see e.g., \cite{DimSjo99}), we finally
obtain
\begin{equation}  \label{eq:mean-beh-boundary-husimis}
\mathcal{H}_k(q,p) \equiv \frac{1}{N(k)}\sum_{k_n\leq k}h_n(q,p)
=\frac{2}{A \pi}\, \sqrt{1-p^2} + \Osk \,\,  ,
\end{equation}

In the derivation of \eqref{eq:mean-beh} from \eqref{eq:mean-int} we have 
assumed that $|p|<1$ because then the stationary points are 
non--degenerate. For $|p|>1$ the stationary points become complex 
and the integral is  exponentially decreasing for $k\to\infty$. 

Previously, such a $\sqrt{1-p^2}$
behaviour appeared in
the context of Fredholm methods for Poincar\'e Husimi functions
\cite{SimSar2000}
and was also obtained in connection 
with the inverse participation ratio \cite{BieKapHagHel2001}.

\BILD{t}
  {
  \begin{center}
    \PSImagx{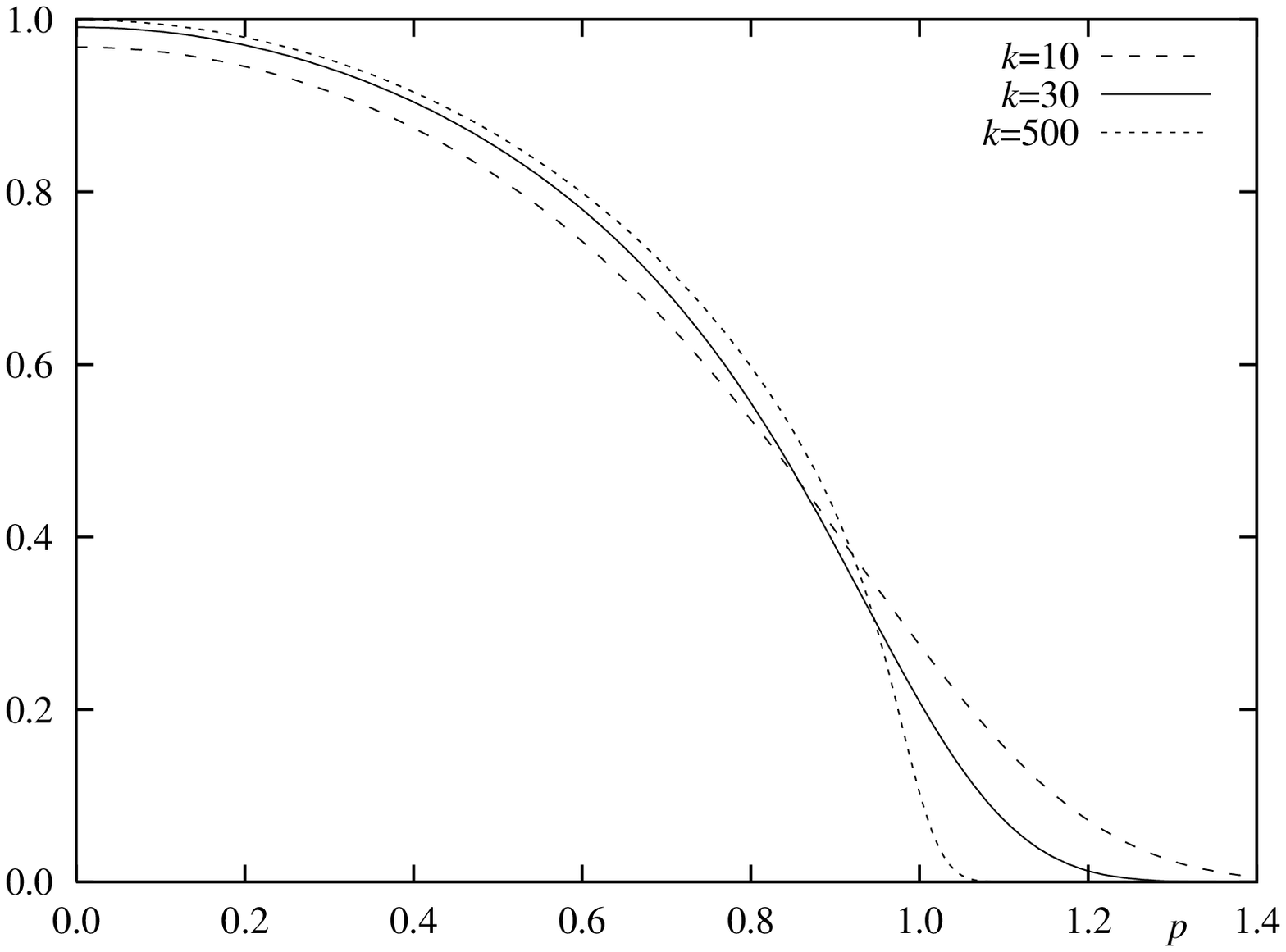}{14cm}
   \end{center}
  }
  {Comparison of the uniformized asymptotic behaviour
   $F_k(p)$, see \eqref{eq:uniform}, with $\frac{|\tilde b|^2}{\Im b}=1$ 
   and for 
  $k=10,30,500$. The asymptotic semi--circle behaviour is reached slowly.}
  {fig:uniform-circle}

Next we want to derive a uniform approximation which describes the 
mean behaviour of the Husimi functions near $|p|=1$ and the crossover 
from the regime $|p|<1$ to the exponential decrease for $|p|>1$. 
We will study the case $p\approx 1$, the case $p\approx -1$ is 
completely analogous. Let $\varphi_0$ be the angle corresponding
to the direction of
$\ta (q)$ and expanding  the amplitude and phase function in 
\eqref{eq:mean-int} around $\varphi_0$ leads to
\begin{equation}
\begin{split}
 \sum_{n\in \N}&\rho(k-k_n)h_n(q,p)
=\frac{4k^2}{(2\pi)^3}\Bigl(\frac{4\pi}{k}\Bigr)^{1/2}
\Int_0^{\infty}\frac{(\Im b)^{1/2}}{|\tilde{b}|}\, \varphi^2
\ue^{-k\frac{\Im b}{|\tilde{b}|^2}(p-1+\varphi^2)^2}\,\, \ud \varphi\; 
\OneOsk\\
&=\frac{4k^2}{(2\pi)^3}\Bigl(\frac{4\pi}{k}\Bigr)^{\frac{1}{2}}
\Int_0^{\infty}\frac{(\Im b)^{1/2}}{|\tilde{b}|}\, x^{1/2}
\ue^{-k\frac{\Im b}{|\tilde{b}|^2}(p-1+x)^2}\,\, \ud x\; \OneOsk\\
&=\ue^{-k\frac{\Im b}{|\tilde{b}|^2}(p-1)^2}
\frac{(2k)^{3/4}}{2\pi^{5/2}}\left(\frac{|\tilde{b}|^2}{\Im b}\right)^{1/4}
\Int_0^{\infty} x^{1/2}
\ue^{-\frac{(2k\Im b)^{1/2}}{|\tilde{b}|}(p-1)x-\frac{x^2}{2}}\,\, 
\ud x\; \OneOsk\\
&=\frac{(2k)^{3/4}}{(2\pi)^2}\, \ue^{-k\frac{\Im b}{2|\tilde{b}|^2}(1-p)^2}
\left(\frac{|\tilde{b}|^2}{\Im b}\right)^{1/4} 
    D_{-3/2}\bigg(\frac{(2k\Im b)^{1/2}}{|\tilde{b}|}\, (p-1)\bigg)\; 
     \OneOsk\, ,
\end{split}
\end{equation}
where $D_{-3/2}(x)$ denotes the parabolic cylinder function and we have used 
one of the standard integral representations, see e.g.~\cite{AbrSte84}.

This result was derived under the assumption $p\approx 1$ 
such that $(p^2-1)\approx 2(p-1)$. Substituting
$(p-1)$ by $(p^2-1)/2$ allows to combine the results for the different 
$p$--regions in one formula 
\begin{equation}\label{eq:uniform-mean}
\sum_{n\in \N}\rho(k-k_n) h_n(q,p)=\frac{k}{\pi^2}\,  F_k(p) \,\OneOsk\, , 
\end{equation}
where 
\begin{equation}\label{eq:uniform} 
F_k(p)=\frac{1}{2(2k)^{1/4}}\ue^{-k\frac{\Im b}{8|\tilde{b}|^2}(1-p^2)^2}
\left(\frac{|\tilde{b}|^2}{\Im b}\right)^{1/4} 
    D_{-3/2}\bigg(\frac{(k\Im b)^{1/2}}{2^{1/2}|\tilde{b}|}\, (p^2-1)\bigg)\; 
\, .
\end{equation}
For $|p|<1$ one has $F_k(p)=\sqrt{1-p^2}+O(k^{-1})$, 
since $D_{-3/2}(x)\sim 2^{3/2}|x|^{1/2}
\ue^{x^2/4}$ for $x\to -\infty$.  
Recall, that $\tilde{b}$ is defined in equation~\eqref{eq:tilde-b-def}.
In figure \ref{fig:uniform-circle} 
we compare the expression \eqref{eq:uniform} with 
$|\tilde b|^2/\Im b=1$ 
for different values of $k$. 
It is clearly visible that the asymptotic result is reached 
slowly with increasing $k$.

Integrating \eqref{eq:uniform}, analogous to the transition
from \eqref{eq:mean-beh} to \eqref{eq:mean-beh-boundary-husimis},
one can compare the uniformized mean behaviour with the numerical result.
In figure~\ref{fig:uniform-circle-with-data}  a section of $\cH_k(q,p)$
at $q=3.0$ is shown for $k=125$, compare with figure~\ref{fig:sandburgen}a).
The remaining differences are due to higher order corrections.

\BILD{b}
  {
  \begin{center}
    \PSImagx{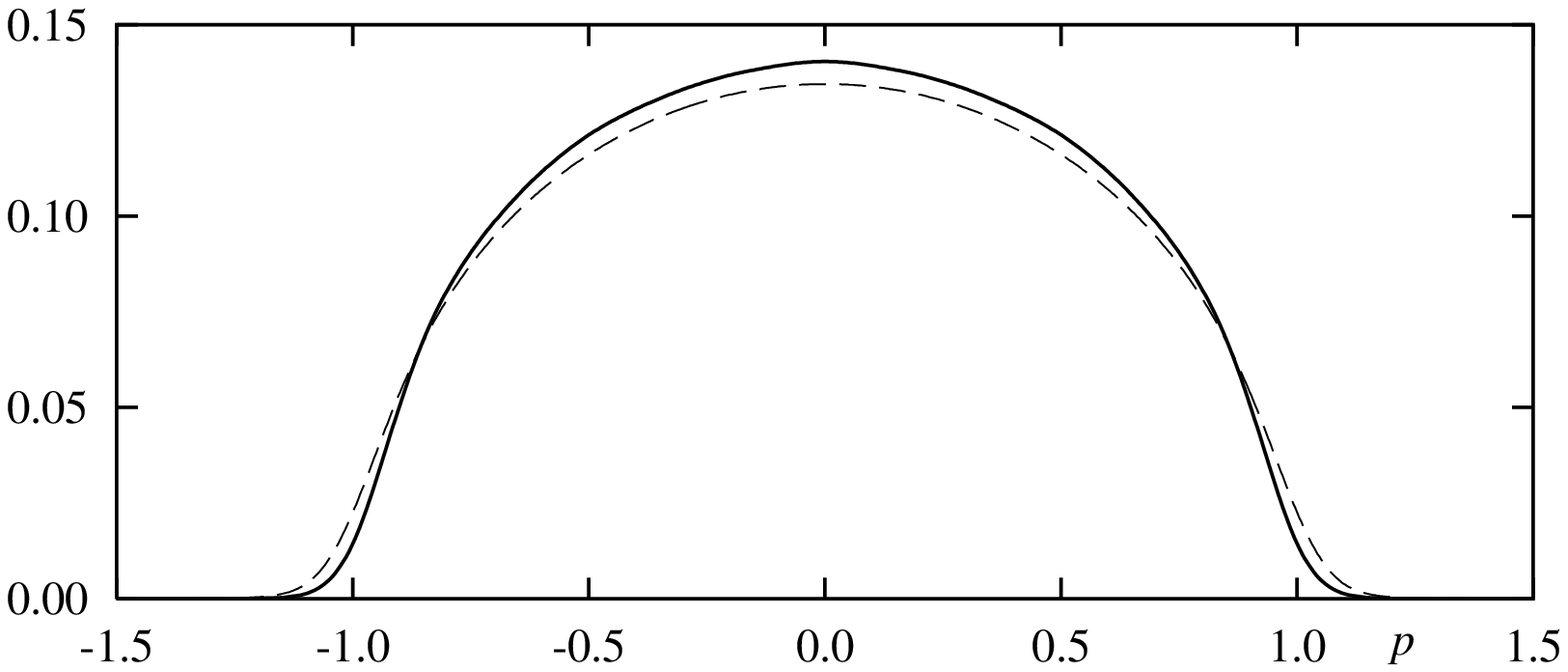}{14cm}
  \end{center}
  }
  {The full curve shows a section of $\cH_k(q,p)$ at $q=3.0$
   with $k=125$ for the desymmetrized \limacon billiard, 
   see figure~\ref{fig:sandburgen}a),
   and the second line is the uniformized mean behaviour.
   The remaining deviations are caused by higher order corrections.}
  {fig:uniform-circle-with-data}

In the derivation of the results \eqref{eq:mean-beh} and  
\eqref{eq:uniform-mean} we have implicitly assumed that the boundary 
of $\Omega$ is sufficiently smooth, because only then we can use 
the stationary phase formula.  But it is easy to extend the results 
to the case that the boundary is only piecewise smooth. Since we multiply in 
\eqref{eq:mult-by-coh-states} by a coherent state centered in 
$q$, all the following computations remain valid if 
$q$ is in the smooth part of the boundary, since the contributions from the 
singular points are exponentially suppressed then. 
So it could only happen that some additional mass sits at the 
singular points of the boundary, i.e.\ that we have 
\begin{equation}
\lim_{k\to\infty}\frac{1}{N(k)}\sum_{k_n\leq k}h_n(q,p)
=\frac{2}{A \pi}\, \sqrt{1-p^2} + \mu_S(p,q)\,\,  ,
\end{equation}
where $\mu_S(p,q)\, \ud p\,\ud q$ is a measure supported on the singular 
part of the boundary. 
Since $h_n(q,p)$ is positive, and $\frac{2}{A \pi}\, \sqrt{1-p^2}$ 
is absolutely 
continuous with respect to the Lebesgue measure, we have $\mu_S\geq 0$.
\footnote{To see this, 
assume that $\mu_S$ has a negative part, then there exist a 
point $z_0=(p_0,q_0)$ and constants 
$\varepsilon_0,C>0$ such that 
$\iint_{|z-z_0|\leq \varepsilon} \mu_S\, \ud z\leq -C$ for all 
$\varepsilon\leq \varepsilon_0$. This implies 
$\lim_{\varepsilon\to 0} \iint_{|z-z_0|\leq \varepsilon} 
\lim_{k\to\infty}\frac{1}{N(k)}\sum_{k_n\leq k}h_n(z)\, \ud z \leq -C$, 
which contradicts the positivity of the $h_n(z)$.}
Now the completeness relation gives
$\lim_{k_n\to\infty}\frac{1}{2}
\iint \la\n(q),\x(q)\ra h_n(q,p)\, \ud p\ud q=1$ 
and $\frac{1}{2}\iint \la\n(q),\x(q)\ra\frac{2}{A \pi}\, 
\sqrt{1-p^2}\, \ud p\ud q=1$ and therefore
\begin{equation}
\Int_{-1}^{1}\Int_{\pa\Omega} \la\n(q),\x(q)\ra \mu_S(p,q)\, 
\ud q\,\ud p =0\,\, .
\end{equation}
But for a star-shaped billiard one can choose the origin of the coordinate 
system such that $ \la\n(q),\x(q)\ra >0$ for all $q\in \pa\Omega$, and so 
$\mu_S=0$. 
 Therefore  \eqref{eq:mean-beh} and  
\eqref{eq:uniform-mean} remain true for star-shaped billiards with piecewise 
smooth boundary with the only possible modification 
that the error term might decay more slowly at the singular points of the 
boundary.

\section{From Husimi functions in phase space to Husimi functions 
         on the boundary} \label{sec:Poincare-Husimi-vs-Husimi}

In this section we derive a direct relation between the 
Husimi function in phase space and the one on the Poincar\'e section,
as given by eq.~\eqref{eq:def-p-husimi-0}.
By this we obtain a physical interpretation 
of the Poincar\'e Husimi representation. 
For the calculations in this section we have to assume 
that the billiard domain $\Omega$ is convex. 
Let $\psi$ be a solution of the Helmholtz equation \eqref{eq:Helmholtz} 
in $\Omega$ which satisfies Dirichlet boundary condition on $\pa\Omega$. 
Any such function can be represented as 
\begin{equation}\label{eq:u-to-psi}
\psi(\x)=-\Int_{\pa\Omega} G_k(\x-\x(s))u(s)\,\, \ud s
\end{equation}
where $G_k(\x-\y)$ is a free Green function and $u(s)$ is the normal 
derivative of $\psi$ on the boundary. 

Let $\psi_{\z}$ be a coherent state \eqref{eq:def-of-int-cs} 
centered at $\z=(\p,\q)\in T^*\R^2$, 
for reasons of simplicity 
we restrict
ourselves to the case of a non--squeezed symmetrical state, 
i.e. $B=\ui\,\bigl(\begin{smallmatrix} 1&0\\ 0&1\end{smallmatrix}\bigr)$,
and omit the index $B$ in the following.
We want to compute the overlap $\la\psi, \psi_{\z}\ra $, given by
\begin{equation}\label{eq:proj-greens}
\la \psi,\psi_{\z}\ra_{\Omega} 
  =-\Int_{\pa\Omega} \la G_k(\cdot-\x(s)),\psi_{\z}\ra_{\Omega} 
   \Cu(s)\,\, \ud s \;\;,
\end{equation}
and we observe that
\begin{equation}\label{eq:scalar-prod}
\la G_k(\cdot-\x(s)),\psi_{\z}\ra_{\Omega}  
={\mathbf G}_k^{\dagger}\psi_{\z}(\x(s))    
\end{equation}
where 
\begin{equation}
{\mathbf G}_k=\lim_{\varepsilon\to 0}\frac{-1}{\Delta+k^2+\ui\varepsilon}
\end{equation}
is the resolvent operator, whose kernel is the Green function. From equation
\eqref{eq:scalar-prod} we see that the function 
${\mathbf G}_k^{\dagger}\psi_{\z}$ is restricted to the billiard boundary.
For the resolvent operator we use the integral representation
\begin{equation}\label{eq:integral-rep}
\mathbf{G}_k^{\dagger}
=\frac{\ui}{k}\Int_{-\infty}^0\ue^{\ui kt}\,\mathbf{U}(t)\,\, \ud t
\end{equation}
where $\mathbf{U}(t)=\ue^{\frac{\ui}{k}t\Delta}$ is the 
free time evolution operator with $1/k$ playing the role of $\hbar$, and 
inserting equation \eqref{eq:integral-rep} into \eqref{eq:scalar-prod} 
we obtain
\begin{equation}
\la G_k(\cdot-\x(s)),\psi_{\z}\ra_{\Omega}  
=\frac{\ui}{k}\Int_{-\infty}^0\ue^{\ui kt}\,\mathbf{U}(t)
\psi_{\z}(\x(s))\,\, \ud t\;\;.              
\end{equation}

But the free time evolution of a coherent state centered in $\z$
is well known (see e.g.~\cite{Hag80,Lit86})
to give again a coherent state, centered around the 
image of $\z$ under the  classical flow and 
with transformed variance,  
\begin{equation}
\mathbf{U}(t)\psi_{\z}(\x)
  =\ue^{\ui k \p^2 t} \left(\frac{k}{\pi}\right)^{1/2}
\frac{1}{1+2\ui t}\, 
  \ue^{\ui k[\la \p, \x-\q(t)\ra+\frac{\ui}{2(1+2\ui t)}(\x-\q(t))^2]}\,\, ,
\end{equation}
with $\q(t)=\q+2t\p$.  Therefore, 
${\mathbf G}_k^{\dagger}\psi_{\z}(\x)$ has the structure of
 a Gaussian beam 
emanating from the point $\q $ in direction $\p$ backwards in time. 
If we introduce a new coordinate system $\x=(x_{||},x_{\perp})$ centered at 
$\q$ with $x_{||}$ parallel to $\p$ and $x_{\perp}$ perpendicular to 
$\p$, we obtain by a stationary phase approximation
that for $x_{\perp}$ and $1-|\p|$ small (i.e.\ near the energy shell)
\begin{equation}\label{eq:gaussian-beam}
{\mathbf G}_k^{\dagger}\psi_{\z}(\x)=
\frac{\ui}{\sqrt{2}k(1+\ui x_{||} )^{1/2}}
\,\ue^{\ui k[x_{||}+\frac{\ui}{2(1+\ui x_{||})}x_{\perp}^2
+\frac{\ui}{2} (1-|\p|)^2]}  \; \OneOsk
\end{equation}
holds, where  we have assumed that $x_{||}<0$. For $x_{||}\approx 0$ 
and $x_{||}> 0$ 
the integral leads to an error function which describes the transition 
from the  exponentially decaying regime with $x_{||}> 0$ to the regime 
$x_{||}<0$ in \eqref{eq:gaussian-beam}. For $|\p|=1$ the result reads 
\begin{equation}
\label{eq:GaussianBeam}
\begin{split}
{\mathbf G}_k^{\dagger}\psi_{\z}(\x)\Bigl\lvert_{|\p|=1}
=\frac{\ui}{\sqrt{2} k(1+\ui x_{||} )^{1/2}}
&\,\ue^{\ui k[x_{||}+\frac{\ui}{2(1+\ui x_{||})}x_{\perp}^2]}\\
&\times\frac{1}{2}\erfc\bigg(\sqrt{\frac{k}{2}}\,
             \frac{x_{||}}{(1+\ui x_{||} )^{1/2}}\bigg) \; \OneOsk\,\, ,
\end{split}
\end{equation}
where $\erfc(z)$ denotes the complementary error function, 
and the absolute value of this expression  is shown in 
figure \ref{fig:gaussian-beam}.

\BILD{tbh}
     {
\begin{center}
    \PSImagx{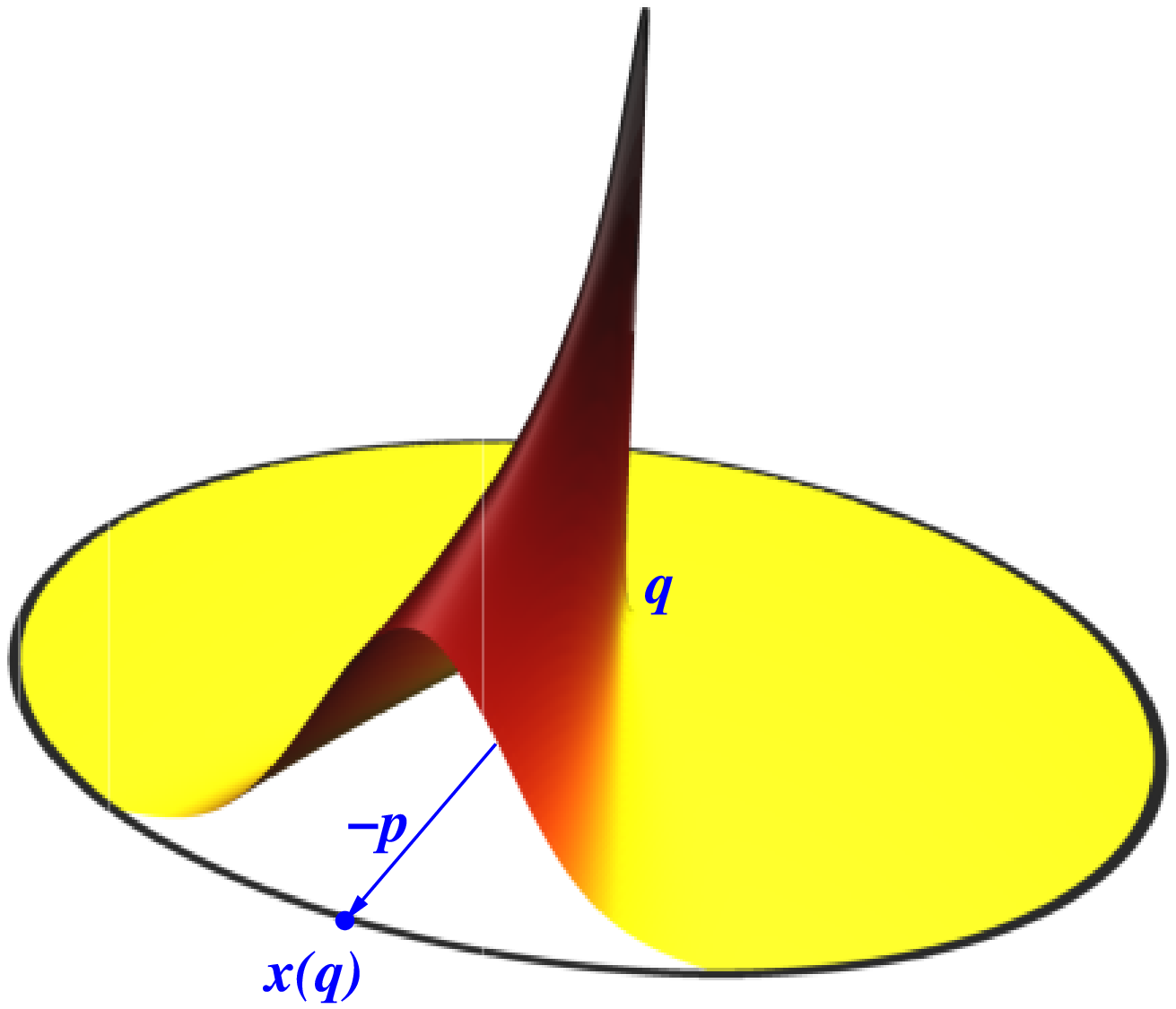}{14cm}
\end{center}     }
     {Illustration of a Gaussian beam as given by \eqref{eq:GaussianBeam}
      inside the \limacon billiard at $\varepsilon=0.3$.}
{fig:gaussian-beam}

Next we want to evaluate this expression 
on the boundary. To this end,
let $\x(q)$ be the point of intersection between the boundary and 
the line from $\q$ in direction $-\p$. 
(Here we need the assumption that the billiard domain $\Omega$ 
is convex, in order that there 
is only one such point.)
Then we obtain with
$\x(s)=\x(q)+\ta (q) (s-q)-\frac{\kappa(q)}{2}\, \n(q)(s-q)^2+O((s-q)^3)$
that 
\begin{align}
x_{||}&=|\q-\x(q)|+p(s-q)-\frac{\kappa(q)}{2}
\, (1-p^2)^{1/2}(s-q)^2+O((s-q)^3)\\
x_{\perp}&=(1-p^2)^{1/2}(s-q)+O((s-q)^2)
\end{align}
where $p:=\la\hat{\p},\ta\ra\in [-1,1]$. Inserting these expressions in 
\eqref{eq:gaussian-beam} gives 
\begin{equation}\label{eq:restr-gauss-beam}
\la G_k(\cdot-\x(s)),\psi_{\z}\ra_{\Omega}=
\frac{\ui \pi^{1/4}}{\sqrt{2}k^{5/4}}\,
\frac{1}{(1-p^2)^{1/4}}
\ue^{\ui k|\q-\x(q)|+\ui \theta} \ue^{-\frac{k}{2}(1-|\p|)^2} \cs(s)
\; \OneOsk \;\;,
\end{equation}
where $\cs(s)$ is a coherent state on the boundary, as defined in 
\eqref{eq:def-of-bound-cs}, with variance 
$b=\frac{\ui(1-p^2)}{1+\ui |\q-\x(q)|}-\kappa(q)(1-p^2)^{1/2}$,  
and $\ue^{\ui\theta}=\frac{(|\q-\x(q)|+\ui )^{1/2}}{(|\q-\x(q)|^2+1)^{1/4}}$.  
Notice that although we started with a symmetric coherent  state in the 
interior, the projected coherent state on the boundary is no longer
 symmetric and 
has a non--trivial squeezing parameter $b$ which depends on the position of 
the original state, the angle of intersection of the ray in direction
$-\p$
with the boundary and the curvature of the boundary.

If we insert the expression \eqref{eq:restr-gauss-beam} into 
\eqref{eq:proj-greens} we  obtain a semiclassical relation between the 
projection of an eigenstate onto a coherent state in the interior and 
the projection of the normal derivative on the boundary onto 
a coherent state on the boundary, 
\begin{equation}
\begin{split}
\la \psi_n, \psi_{\z}\ra_{\Omega} =-\frac{\ui \pi^{1/4}}{\sqrt{2}\,k_n^{5/4}}
\,\frac{1}{(1-p^2)^{1/4}}
& \ue^{\ui k_n|\q-\x(q)|+\ui \theta}
\,\ue^{-\frac{k_n}{2}(1-|\p|)^2} \\
& \times\la u_n, \csn\ra_{\pa\Omega}
\;\OneOskn\,\, .
\end{split}
\end{equation}
In turn from this we obtain the central result of this section,
a direct relation between the corresponding 
Husimi functions 
\begin{equation}\label{eq:H-h-relation}
H_n(\p,\q)=\delta_{k_n}(1-|\p|)\, \frac{1}{4}\,
\frac{ h_n(q,p)}{\sqrt{1-p^2}}\;\OneOskn\,\, ,
\end{equation}
with
\begin{equation}
\delta_{k_n}(1-|\p|):= \left(\frac{k_n}{\pi}\right)^{1/2}
\ue^{-k_n (1-|\p|)^2}\,\, .
\end{equation}

Let us first discuss the meaning of the individual 
terms on the right hand side of equation \eqref{eq:H-h-relation}. The 
function $\delta_{k_n}(1-|\p|)$ is a delta sequence 
for $k_{n}\to\infty$, and describes 
the localization of $H_n(\p,\q)$ around the energy shell. 
The factor $1/\sqrt{1-p^2}$ 
comes from the projection of the Gaussian beam to the plane 
tangent to the boundary, see figure \ref{fig:gaussian-beam}.

As in the previous section we have assumed that the boundary is 
smooth. But by the localization of the coherent states the results can be 
again extended to the case that the boundary is piecewise smooth, then 
\eqref{eq:H-h-relation} remains valid if $q$ is not a singular point 
of the boundary.

The direct connection between the 
Husimi function in the interior and the 
one on the boundary, given by equation \eqref{eq:H-h-relation},
allows to derive interesting relations between the two Husimi functions
 and can be used to give a direct physical interpretation of 
the Husimi function on the boundary. 
From equation~\eqref{eq:int-prob-distr} together with 
relation \eqref{eq:H-h-relation} we obtain 
\begin{equation} \label{eq:innen-aussen}
\la \psi_n, \bfA\psi_n\ra_{\Omega} =\Int_{-1}^{1}\Int_{\pa\Omega} 
\frac{h_n(q,p)}{4\sqrt{1-p^2}}\,  \langle a\rangle(q,p)l(q,p)\, 
\ud q\,\ud p +O(k_n^{-1/2})\,\, ,
\end{equation}
where $l(q,p)$ denotes the length of a ray emanating from $\q(q)\in\pa\Omega$ 
in the direction determined by $p$ until it hits the boundary again. 
Furthermore,
\begin{equation}
\langle a\rangle(q,p):= \frac{1}{l(q,p)}\Int_0^{l(q,p)}
a(\q(q)+t\e(q,p),\e(q,p))\, \ud t \;\;,
\end{equation}
is the mean value of the classical observable between two bounces,
where $\e(q,p)$ denotes the unit vector at $\q(q)$ in direction $p$.
A relation of the same type as \eqref{eq:innen-aussen}
has been obtained recently by different methods in \cite{LeeCre2003:p}
for certain localized functions on the boundary.

We conclude from  relation \eqref{eq:innen-aussen} that 
\begin{equation}\label{eq:weighted-hus}
\frakh_n(q,p):=\frac{1}{4}\,\frac{h_n(q,p)}{\sqrt{1-p^2}}
\end{equation}
is a reduction of the probability density defined by 
the Husimi function 
on the whole phase space to the boundary. 
So if one wants a proper representation of eigenfunctions on the Poincar{\'e}
section which is an approximate probability density, and whose general 
properties are independent of the billiard shape, then 
\eqref{eq:weighted-hus} seems to be the best choice.   
Of course a drawback of 
the function \eqref{eq:weighted-hus} is the
singularity of $1/\sqrt{1-p^2}$ at $p=\pm 1$ 
which is relevant at any finite energy.
So for numerical computations the definition \eqref{eq:def-p-husimi-0}
is more suitable and the importance of \eqref{eq:weighted-hus}
lies in the physical interpretation.

In particular, relation \eqref{eq:innen-aussen}
implies  an 
asymptotic normalization condition on $\frakh_n(q,p)$, 
\begin{equation}
\Int_{-1}^{1}\Int_{\pa\Omega} 
\frakh_n(q,p)\, l(q,p)\,\, 
\ud q\,\ud p  =1+O(k_n^{-1/2})\,\, .
\end{equation}
Since $l(q,p)\ud q\,\ud p$ is the phase space volume in the energy 
shell corresponding
to the volume element $\ud q\,\ud p$ of the Poincar\'e section, the factor 
$l(q,p)$ can be viewed as a normalization  which 
makes $\frakh_n(q,p)$ independent of the billiard shape, i.e.\ 
for any $\cD\subset \pa \Omega\times [-1,1]$, we get 
that $\int_{\cD}\frakh_n(q,p) l(q,p) \ud q\,\ud p$ 
is the probability for the particle in the state $\psi_n$ to 
be found in the region 
$\hat{D}:=\Pi^{-1}\cD$ on the energy shell, where the map $\Pi$ 
describes the projection of the domain $\hat{D}$ to the boundary.

We would like to close this section with some remarks on the implications 
of quantum ergodicity to the behaviour of the Poincar\'e
Husimi functions. If the classical 
billiard flow in $\Omega$ is ergodic, then the quantum ergodicity theorem 
\cite{GerLei93, ZelZwo96} (see \cite{BaeSchSti98} for an introduction)
tells us that almost all Husimi functions 
$H_n(\p,\q)$ tend weakly to $\frac{1}{2\pi \vol(\Omega)}$. 
Our result \eqref{eq:H-h-relation}
then immediately implies that in the semiclassical limit
almost all Poincar\'e Husimi functions $h_n(q,p)$
tend to $\frac{2}{\pi\vol(\Omega)}\, \sqrt{1-p^2}$ in the weak sense. 
So this proves a quantum ergodicity theorem for the boundary Husimi functions. 
Recently related results 
have been obtained establishing quantum ergodicity 
for observables on the Poincar\'e section
\cite{GerLei93,HasZel2002:p,Bur2003:p}.
Notice that the $\sqrt{1-p^2}$ behaviour is also visible in the plot
of $h_n(q,p)$ for the irregular state shown in 
fig.~\ref{fig:husimi-example}c) for the ergodic cardioid billiard.

\section{Summary}

Poincar\'e representations of eigenstates play
an important role in several areas.
However, a priori there is no unique way for their definition.
In this paper we single out the definition given by 
\eqref{eq:def-p-husimi-0} and
show that the asymptotic mean behaviour of 
these Husimi functions is proportional to $\sqrt{1-p^2}$.
For this asymptotic semi-circle behaviour we in addition 
derive a uniform asymptotic formula.
Furthermore we establish a direct relation between the Husimi function
in phase space and the Poincar\'e Husimi function \eqref{eq:def-p-husimi-0} 
on the billiard boundary.
By this a physically meaningful interpretation, 
see equation~\eqref{eq:H-h-relation},
of the previously ad--hoc chosen
definition for the Poincar\'e Husimi function is obtained.
 Namely, the Poincar\'e Husimi function $\frakh_n(q,p)$ can be
viewed as a probability density on the Poincar\'e section.
For ergodic systems our result implies a 
quantum ergodicity theorem for the Poincar\'e Husimi functions,
i.e.~almost all Poincar\'e Husimi functions become
equidistributed with respect to the appropriate measure.

\vspace*{0.5cm}

\noindent {\bf Acknowledgment}

AB and RS would like to thank the 
Mathematical Sciences Research Institute, Berkeley, USA,
for financial support and
hospitality where part of this work was done.



\begin{thebibliography}{10}

\bibitem{Sto99}
H.-J. St{\"o}ckmann: {\em Quantum chaos\/}, Cambridge University Press,
  Cambridge,  (1999).

\bibitem{Wig32}
E.~P. Wigner: {\em On the quantum correction for thermodynamic equilibrium\/},
  Phys. Rev. {\bf 40} (1932) ~749--759.

\bibitem{Hus40}
K.~Husimi: {\em Some formal properties of the density matrix\/}, Proc. Phys.
  Math. Soc. Jpn. {\bf 22} (1940) ~264--314.

\bibitem{BaeSch99}
A.~B\"acker and R.~Schubert: {\em Chaotic eigenfunctions in momentum space\/},
  J. Phys. A {\bf 32} (1999) ~4795--4815.

\bibitem{VebRobLiu99}
G.~Veble, M.~Robnik and J.~Liu: {\em Study of regular and irregular states in
  generic systems\/}, J. Phys. A {\bf 32} (1999) ~6423--6444.

\bibitem{CrePerCha93}
B.~Crespi, G.~Perez and S.-J. Chang: {\em Quantum Poincar\'e sections for
  two-dimensional billiards\/}, Phys. Rev. E {\bf 47} (1993) ~986--991.

\bibitem{TuaVor95}
J.~M. Tualle and A.~Voros: {\em Normal modes of billiards portrayed in the
  stellar (or nodal) representation\/}, Chaos, Solitons and Fractals {\bf 5}
  (1995) ~1085--1102.

\bibitem{SimVerSar97}
F.~P. Simonotti, E.~Vergini and M.~Saraceno: {\em Quantitative study of scars
  in the boundary section of the stadium billiard\/}, Phys. Rev. E {\bf 56}
  (1997) ~3859--3867.

\bibitem{BieKapHagHel2001}
W.~E. Bies, L.~Kaplan, M.~R. Haggerty and E.~J. Heller: {\em Localization of
  eigenfunctions in the stadium billiard\/}, Phys. Rev. E {\bf 63} (2001)
  ~066214.

\bibitem{Kap2000}
L.~Kaplan: {\em Periodic orbit effects on conductance peak heights in a chaotic
  quantum dot\/}, Phys. Rev. E {\bf 62} (2000) ~3476--3488.

\bibitem{BaeManHucKet2002}
A.~B\"acker, A.~Manze, B.~Huckestein and R.~Ketzmerick: {\em Isolated
  resonances in conductance fluctuations and hierarchical states\/}, Phys. Rev.
  E {\bf 66} (2002) ~016211 (8 pages).

\bibitem{BaeSch2002a}
A.~B\"acker and R.~Schubert: {\em Amplitude distribution of eigenfunctions in
  mixed systems\/}, J. Phys. A {\bf 35} (2002) ~527--538.

\bibitem{BaeSch2002b}
A.~B\"acker and R.~Schubert: {\em Autocorrelation function of eigenstates in
  chaotic and mixed systems\/}, J. Phys. A {\bf 35} (2002) ~539--564.

\bibitem{KlaSmi96}
D.~Klakow and U.~Smilansky: {\em Wavefunctions, expectation values and scars on
  Poincar\'e sections -- A scattering approach\/}, J. Phys. A {\bf 29} (1996)
  ~3213--3231.

\bibitem{FriDor97}
S.~D. Frischat and E.~Doron: {\em Quantum phase-space structures in classically
  mixed systems: a scattering approach\/}, J. Phys. A {\bf 30} (1997)
  ~3613--3634.

\bibitem{HenSchSch2003}
M.~Hentschel, H.~Schomerus and R.~Schubert: {\em Husimi functions at dielectric
  interfaces: Inside-outside duality for optical systems and beyond\/},
  Europhys. Lett. {\bf 62} (2003) ~636--642.

\bibitem{Bog92}
E.~B. Bogomolny: {\em Semiclassical quantization of multidimensional
  systems\/}, Nonlinearity {\bf 5} (1992) ~805--866.

\bibitem{Pau97}
T.~Paul: {\em Semi-classical methods with emphasis on coherent states\/}, in:
  {\em Quasiclassical methods (Minneapolis, MN, 1995)\/}, vol.~95 of {\em IMA
  Vol. Math. Appl.\/},  51--88, Springer, New York,  (1997).

\bibitem{BaeFueSchSte2002}
A.~B\"acker, S.~F\"urstberger, R.~Schubert and F.~Steiner: {\em Behaviour of
  boundary functions for quantum billiards\/}, J. Phys. A {\bf 35} (2002)
  ~10293--10310.

\bibitem{BaeSchSti98}
A.~B\"acker, R.~Schubert and P.~Stifter: {\em Rate of quantum ergodicity in
  Euclidean billiards\/}, Phys. Rev. E {\bf 57} (1998) ~5425--5447; erratum
  ibid. {\bf 58} (1998) 5192.

\bibitem{Rel1940}
F.~Rellich: {\em Darstellung der {E}igenwerte von $\Delta u+\lambda u=0$ durch
  ein {R}andintegral\/}, Math. Z. {\bf 46} (1940) ~635--636.

\bibitem{BerWil84}
M.~V. Berry and M.~Wilkinson: {\em Diabolical points in the spectra of
  triangles\/}, Proc. R. Soc. London Ser. A {\bf 392} (1984) ~15--43.

\bibitem{Boa94}
P.~A. Boasman: {\em Semiclassical accuracy for billiards\/}, Nonlinearity {\bf
  7} (1994) ~485--537.

\bibitem{Rob83}
M.~Robnik: {\em Classical dynamics of a family of billiards with analytic
  boundaries\/}, J. Phys. A {\bf 16} (1983) ~3971--3986.

\bibitem{Rob84}
M.~Robnik: {\em Quantising a generic family of billiards with analytic
  boundaries\/}, J. Phys. A {\bf 17} (1984) ~1049--1074.

\bibitem{DimSjo99}
M.~Dimassi and J.~Sj{\"o}strand: {\em Spectral Asymptotics in the
  Semi-Classical Limit\/}, Cambridge University Press, Cambridge,  (1999).

\bibitem{SimSar2000}
F.~P. Simonotti and M.~Saraceno: {\em Fredholm methods for billiard
  eigenfunctions in the coherent state representation\/}, Phys. Rev. E {\bf 61}
  (2000) ~6527--6537.

\bibitem{AbrSte84}
M.~Abramowitz and {I. A. Stegun (eds.)}: {\em Pocketbook of Mathematical
  Functions\/}, Verlag Harri Deutsch, Thun -- Frankfurt/Main, abridged edn.,
  (1984).

\bibitem{Hag80}
G.~A. Hagedorn: {\em Semiclassical quantum mechanics. {I}. {T}he {$\hbar
  \rightarrow 0$} limit for coherent states\/}, Comm. Math. Phys. {\bf 71}
  (1980) 1 ~77--93.

\bibitem{Lit86}
R.~G. Littlejohn: {\em The semiclassical evolution of wave packets\/}, Phys.
  Rep. {\bf 138} (1986) 4-5 ~193--291.

\bibitem{LeeCre2003:p}
S.-Y. Lee and S.~C. Creagh: {\em Wavefunction statistics using scar states\/},
  preprint nlin.CD/0304018.

\bibitem{GerLei93}
P.~G\'erard and E.~Leichtnam: {\em Ergodic properties of eigenfunctions for the
  Dirichlet problem\/}, Duke Math. J. {\bf 71} (1993) ~559--607.

\bibitem{ZelZwo96}
S.~Zelditch and M.~Zworski: {\em Ergodicity of eigenfunctions for ergodic
  billiards\/}, Commun. Math. Phys. {\bf 175} (1996) ~673--682.

\bibitem{HasZel2002:p}
A.~Hassell and S.~Zelditch: {\em Quantum ergodicity of boundary values of
  eigenfunctions\/}, preprint math.SP/0211140  (2002).

\bibitem{Bur2003:p}
N.~Burq: {\em Quantum ergodicity of boundary values of eigenfunctions: A
  control theory approach\/}, preprint math.AP/0301349  (2003).

\end{thebibliography}
\end{document}